\documentclass[12pt]{article}
\usepackage{amsmath,amsfonts,amssymb,graphicx}
\usepackage{graphics, setspace}

\usepackage{epstopdf}
\DeclareGraphicsExtensions{.eps}

\begin{document}
\thispagestyle{empty}

\def\theequation{\arabic{section}.\arabic{equation}}
\def\a{\alpha}
\def\b{\beta}
\def\g{\gamma}
\def\d{\delta}
\def\dd{\rm d}
\def\e{\epsilon}
\def\ve{\varepsilon}
\def\z{\zeta}
\def\B{\mbox{\bf B}}\def\cp{\mathbb {CP}^3}

\newcommand{\h}{\hspace{0.5cm}}

\begin{titlepage}
\rightline{HU-EP-14/58}
\rightline{HU-Mathematik-2014-36}

\vspace*{1.cm}
\renewcommand{\thefootnote}{\fnsymbol{footnote}}
\begin{center}
{\Large \bf A HHL 3-point correlation function in the $\eta$-deformed $AdS_5\times S^5$}
\end{center}
\vskip 1.2cm \centerline{\bf Changrim  Ahn$^{1,2}$ and Plamen Bozhilov$^{2,3}$}

\vskip 10mm
\centerline{\sl$\ ^1$ Institut f\"ur Mathematik, 
Institut f\"ur Physik und IRIS Adlershof}
\centerline{\sl Humboldt-Universit\"at zu Berlin}
\centerline{\sl Zum Gro\ss en Windkanal 6, 12489 Berlin, Germany}
\vskip .5cm
\centerline{\sl$\ ^2$Department of Physics} \centerline{\sl Ewha Womans
University} \centerline{\sl DaeHyun 11-1, Seoul 120-750, S. Korea}
\vskip .5cm
\centerline{\sl$\ ^3$Institute for Nuclear Research and Nuclear Energy}
\centerline{\sl Bulgarian Academy of Sciences, Bulgaria}

\vspace*{0.6cm} \centerline{\tt ahn@ewha.ac.kr,
bozhilov@inrne.bas.bg}

\vskip 20mm

\baselineskip 18pt

\begin{center}
{\bf Abstract}
\end{center}

We derive the 3-point correlation function between two giant magnons 
heavy string states
and the light dilaton operator with zero momentum 
in the $\eta$-deformed $AdS_5\times S^5$
valid for any $J_1$ and $\eta$ in the semiclassical limit.
We show that this result satisfies a consistency relation
between the 3-point correlation function and 
the conformal dimension of the giant magnon.
We also provide a leading finite $J_1$ correction explicitly.

\end{titlepage}
\newpage
\baselineskip 18pt

\def\nn{\nonumber}
\def\tr{{\rm tr}\,}
\def\p{\partial}
\newcommand{\non}{\nonumber}
\newcommand{\bea}{\begin{eqnarray}}
\newcommand{\eea}{\end{eqnarray}}
\newcommand{\bde}{{\bf e}}
\renewcommand{\thefootnote}{\fnsymbol{footnote}}
\newcommand{\be}{\begin{eqnarray}}
\newcommand{\ee}{\end{eqnarray}}

\vskip 0cm

\renewcommand{\thefootnote}{\arabic{footnote}}
\setcounter{footnote}{0}

\setcounter{equation}{0}
\section{Introduction}
The AdS/CFT duality \cite{AdS/CFT} between string theories on
curved space-times with Anti-de Sitter subspaces and conformal
field theories in different dimensions has been actively
investigated in the last years. A lot of impressive progresses
have been made in this field of research based mainly on the
integrability structures discovered on both sides of the
correspondence (for recent review on the AdS/CFT duality, see
\cite{RO}). For the most studied case of the ${\cal N}=4$ super Yang-Mills theory,
the anomalous dimensions of gauge-invariant single-trace operators
match non-perturbatively with the string energies in the
curved $AdS_{5}\times S^{5}$ background. 
Integrability provides tools to solve the finite-volume spectral problem exactly.

After these successes, one direction of interesting developments is to generalize the duality to larger theories
which include the original AdS/CFT as a special case and the other is to go beyond the spectral problem 
by computing general correlation functions, in particular, the three-point functions, or the structure constants.

An interesting development for the former direction is to study the string theory on the $\eta$-deformed $AdS_5\times S^5$ background \cite{DMV0913}.
The bosonic part of the superstring sigma model Lagrangian on this
$\eta$-deformed background and perturbative worldsheet
$S$-matrix were obtained in \cite{ABF1312}.
The TBA for spectrum and explicit dispersion relation for giant magnon 
\cite{HM06}
have been derived in \cite{ALT0314}.
Finite-size effect on the giant magnon spectrum has been computed in \cite{AP2014}.
For three-point correlation functions, quite a lot of interesting results on both strong and weak coupling regions were 
accumulated although non-perturbative results are much more difficult than the spectral problem.

In this letter, we compute the three-point correlation function of two giant magnon heavy operators with finite-size 
$J_1$ and a single dilaton light operator of the string theory with the $\eta$-deformed $AdS_5\times S^5$
background \cite{DMV0913}. 
Then, we show that this result is consistent with the dispersion relation of the finite-size giant magnon solution
obtained in \cite{AP2014} using Mathematica code.

The paper is organized as follows. 
In Sec. 2, we derive the exact semiclassical structure constant valid for any $J_1$ and $\eta$  
and prove its consistency.
In Sec. 3, we expand it for the case of $J_1\gg T$ ($T$ is the tension of string) and obtain explicit 
expression.
A brief conclusion is in sect.4 and a short Mathematica code for the 
consistency condition is provided in Appendix. 

\setcounter{equation}{0}
\section{Exact semiclassical structure constant}
According to \cite{rt10}, 
the three-point functions of two
"heavy" operators and a "light" operator can be approximated by a
supergravity vertex operator evaluated at the "heavy" classical
string configuration: \bea \nn \langle
V_{H}(x_1)V_{H}(x_2)V_{L}(x_3)\rangle=V_L(x_3)_{\rm classical}. \eea
For $\vert x_1\vert=\vert x_2\vert=1$, $x_3=0$, the correlation
function reduces to \bea \nn \langle
V_{H}(x_1)V_{H}(x_2)V_{L}(0)\rangle=\frac{C_{123}}{\vert
x_1-x_2\vert^{2\Delta_{H}}}. \eea Then, the normalized structure
constant \bea \nn \mathcal{C}_3=\frac{C_{123}}{C_{12}} \eea can be
found from \bea \label{nsc} \mathcal{C}_3=c_{\Delta} V_L(0)_{\rm
classical}, \eea where $c_{\Delta}$ is the normalized constant of the
"light" vertex operator. Actually, we are going to
compute the normalized structure constant
(\ref{nsc}). For the case under consideration, the "light" state is represented by the dilaton
with zero momentum.

According to \cite{Hernandez2}, $C_3$ for the infinite-size giant magnons and dilaton
with zero momentum in the undeformed $AdS_5\times S^5$ is given by
\bea\label{und} C_3&=&c_\Delta^d  \int_{-\infty}^{+\infty} \frac{d\tau_e}{\cosh^4(\kappa \tau_e)}
\int_{-\infty}^{+\infty}d\sigma \left(\kappa^2+\p X_K \bar{\p} X_K\right)
\\ \nn &=& \frac{4 c_\Delta^d}{3 \kappa} \int_{-\infty}^{+\infty}d\sigma \left(\kappa^2+\p X_K \bar{\p} X_K\right),\eea
where $t=\kappa \tau_e$ is the Euclidean $AdS$ time and the term $\p X_K \bar{\p} X_K$
is proportional to the string Lagrangian on $S^2$ computed on the giant magnon solution
living in the $R_t\times S^2$ subspace.

Since here we are interested in {\it finite-size} giant magnons, we have to replace
\bea\nn \int_{-\infty}^{+\infty}d\sigma \to \int_{-L}^{+L}d\sigma
=2\int_{\theta_{min}}^{\theta_{max}}\frac{d\theta}{\theta^{'}},\eea
where $L$ gives the size of the giant magnon and $\theta$ is the non-isometric angle
on the two-sphere \cite{AP2011}.

Going to the $\eta$-deformed $AdS_5\times S^5$ case, we have to compute the term
$\p X_K \bar{\p} X_K$ for this background, which is proportional to the string Lagrangian on $S_{\eta}^{2}$
for {\it finite-size} giant magnons:
\bea\nn L_{S_{\eta}^{2}} = -\frac{T}{2} \p X_K \bar{\p} X_K,\eea
where
$X_K=(\phi_1,\theta)$ are the isometric and non-isometric
string coordinates on $S_{\eta}^{2}$ correspondingly.

Working in conformal gauge and applying the ansatz
\bea\nn \phi_1(\tau,\sigma)= \tau+F_1(\xi),
\h \theta(\tau,\sigma)=\theta(\xi),\h \xi=\alpha\sigma+\beta\tau ,\h \alpha,\beta - constants,\eea one finds
\bea\label{Le1} L_{S_{\eta}^{2}} = -\frac{T}{2} \Bigg\{(\alpha^2-\beta^2) \frac{\theta'^{2}}{1+\tilde{\eta}^2(1-\chi)}
+(1-\chi)\Big[(\alpha^2-\beta^2)(F'_1)^2-2\beta F'_1-1\Big]\Bigg\},\eea
where $\tilde{\eta}$ is related to the deformation parameter $\eta$ according to \cite{ABF1312}
\bea\label{ek} \tilde{\eta}=\frac{2 \eta}{1-\eta^2}, \eea
and new variable $\chi$ is defined by
\bea\nn \chi=\cos^2\theta.\eea
The prime here and below is a derivative $d/d \xi$.
The string tension $T$ for the $\eta$ deformed case is related to the coupling constant $g$ by
\bea\label{T}T=g \sqrt{1+\tilde{\eta}^2}.\eea

The first integrals of the equations of motion $ F'_1$ and $\theta'$
can be written as
\bea\label{fp1} F'_1 = \frac{\beta}{\alpha^2-\beta^2} \left(-\frac{\kappa^2}{1-\chi}+1\right),\eea
\bea\label{tp1} \theta'^{2} = \frac{1+\tilde{\eta}^2(1-\chi)}{(\alpha^2-\beta^2)^2}
\Bigg[(\alpha^2+\beta^2)\kappa^2-\frac{\beta^2\kappa^4}{1-\chi}-\alpha^2(1-\chi)\Bigg].\eea
Inserting (\ref{fp1}), (\ref{tp1}) in (\ref{Le1}), we obtain:
\bea\label{Le2} L_{S_{\eta}^{2}} = -\frac{T}{2}\frac{\beta^2\kappa^2+\alpha^2(\kappa^2-2(1-\chi))}{\alpha^2-\beta^2}.\eea
Now we introduce the new parameters
\bea\nn v=-\frac{\beta}{\alpha}, \h W=\kappa^2,\eea
which leads to
\bea\label{Le3} L_{S_{\eta}^{2}}= -\frac{T}{2} \frac{(1+v^2)W-2(1-\chi)}{1-v^2}.\eea
Therefore, for the case at hand, 
the normalized structure constant takes the form
\bea\label{C0} &&C_3^{\tilde{\eta}}= \frac{8 c_\Delta^d}{3 \sqrt{W}}
\int_{\chi_m}^{\chi_p}\frac{d \chi}{\chi'}
\Bigg[W+ \frac{(1+v^2)W-2(1-\chi)}{1-v^2}\Bigg],\eea
where
\bea\nn \chi_m=\chi_{min},\h \chi_p=\chi_{max}.\eea
One can rewrite Eq.(\ref{tp1}) as
\bea\label{cp} \chi'=\frac{2\tilde{\eta}}{1-v^2}
\sqrt{(\chi_{\eta}-\chi)(\chi_p-\chi)(\chi-\chi_m)\chi},
\eea
where \cite{AP2014} 
\bea
\chi_{m}=1-W,\h \chi_{p}=1-v^2 W,\h \chi_{\eta}=1+\frac{1}{{\tilde\eta}^2}.
\label{param}
\eea
Using this, we can express all the results in terms of $\chi_p,\ \chi_m$ 
by eliminating $v,\ W$. 

Replacing (\ref{cp}) into (\ref{C0}) and using (\ref{param}), we can 
express $C_3^{\tilde{\eta}}$ by 
\bea
C_3^{\tilde{\eta}}=\frac{8c_\Delta^d}{3\tilde{\eta}\sqrt{1-\chi_m}}
\int^{\chi_p}_{\chi_m}\sqrt{\frac{\chi-\chi_m}{(\chi_{\eta}-\chi)
(\chi_p-\chi)\chi}}\ d\chi.
\eea
The integral can be easily expressed by $\mathbf{K}$ and $\mathbf{\Pi}$,
the complete elliptic integrals
of the first and the third kind, respectively, as follows:
\bea\label{C1} C_3^{\tilde{\eta}}= 
\frac{16c_\Delta^d}{3\tilde{\eta}}\frac{\chi_m}{\sqrt{\chi_p(1-\chi_m)
(\chi_\eta-\chi_m)}}
\left[
\mathbf{\Pi}\left(1-\frac{\chi_m}{\chi_p},1-\epsilon\right)
-\mathbf{K}\left(1-\epsilon\right) \right],
\eea
where we introduced a short notation $\epsilon$ by 
\bea
\label{epsil}
\epsilon=\frac{\chi_{m}(\chi_\eta-\chi_p)}{\chi_{p}(\chi_\eta-\chi_m)}.
\eea

Eq.(\ref{C1}) is the main result of this paper, which is an 
{\it exact} semiclassical result for the normalized structure constant 
$C_3^{\tilde{\eta}}$ valid for any
value of $\tilde\eta$ and $J_1$.
Here, $\chi_p$ and $\chi_m$ are determined by the angular momentum $J_1$ and
world-sheet momentum $p$ from the following equations:
\footnote{We express $J_1$ and $p$ in terms of different but equivalent
combinations of elliptic functions compared with 
Eqs. (3.23) and (3.25) in \cite{AP2014}.}
\bea
J_1&=&\frac{2T}{\tilde\eta}
\frac{1}{\sqrt{\chi_p(\chi_\eta-\chi_m)}}
\left[\chi_p\mathbf{K}\left(1-\epsilon\right)
-\chi_m\mathbf{\Pi}\left(1-\frac{\chi_m}{\chi_p},
1-\epsilon\right) \right],\label{J1}
\\
p&=&\frac{2\chi_m}{\tilde\eta}
\sqrt{\frac{1-\chi_p}{\chi_p(1-\chi_m)(\chi_\eta-\chi_m)}}
\left[\mathbf{K}\left(1-\epsilon\right)
-\mathbf{\Pi}\left(\frac{\chi_p-\chi_m}{\chi_p(1-\chi_m)},
1-\epsilon\right)\right].
\eea
The world-sheet energy of the giant magnon is given by
\bea
E=\frac{2T}{\tilde\eta}
\frac{\chi_p-\chi_m}{\sqrt{\chi_p(1-\chi_m)(\chi_\eta-\chi_m)}}\ \mathbf{K}
\left(1-\epsilon\right).\label{E1}
\eea

One of nontrivial check is that the $g$ derivative of 
$\Delta=E-J_1$ should be proportional to 
the normalized structure constant $C_3^{\tilde{\eta}}$ 
since the $g$ derivative of the two-point function inserts 
the dilaton (Lagrangian) operator into the two-point function of 
the heavy operators \cite{Costa}.
This can be expressed by
\bea 
C_3^{\tilde{\eta}}=\frac{8 c_\Delta^d}{3 \sqrt{1+\tilde{\eta}^2}} 
\frac{\p\Delta}{\p g}.\label{check}
\eea
To check that Eqs.(\ref{C1}), (\ref{J1}), and (\ref{E1}) satisfy 
Eq.(\ref{check}), we use the fact that 
\bea
\frac{\p J_1}{\p g}=\frac{\p p}{\p g}=0
\eea
as noticed in \cite{AP2011} for the case of undeformed giant magnon.
From these, we can obtain the expressions for 
$\p\chi_p/\p g$ and $\p\chi_m/\p g$ which can be inserted to 
$\p\Delta/\p g$.
The $\eta$-deformed case involves much more complicated expressions which
can be dealt with the Mathematica. 
In the Appendix, we provide our Mathematica code which confirms
that the structure constant $C_3^{\tilde{\eta}}$ in Eq.(\ref{C1})
do sastisfy the consistency condition (\ref{check}) exactly.

In the limit $\tilde{\eta}\to 0$ with $\tilde{\eta}^2\chi_\eta\to 1$,
Eq.(\ref{C1}) becomes
\bea\label{C10}C_3^{0}=
\frac{16c_\Delta^d}{3}\sqrt{\frac{\chi_p}{1-\chi_m}}
\ \left[
\mathbf{E}\left(1-\epsilon\right)
-\epsilon\mathbf{K}\left(1-\epsilon\right) \right],
\h \epsilon=\frac{\chi_m}{\chi_p}
\eea
where we used an identity $(1-a)\mathbf{\Pi}(a,a)=\mathbf{E}(a)$.
This is the structure constant of the undeformed theory
derived in \cite{AP2011}. 

\setcounter{equation}{0}
\section{Leading finite-size effect on $C_3^{\tilde{\eta}}$}
It is straight forward to compute a leading finite-size effect 
on $C_3^{\tilde{\eta}}$ for $J_1\gg g$ by 
the limit $\epsilon\to 0$ in (\ref{C1}).

First we expand the parameters $\chi_p$, $W$ and $v$ 
for small $\epsilon$ as follows:
\bea\label{chie} &&\chi_{p}= \chi_{p 0}+(\chi_{p 1}+\chi_{p 2}\log\epsilon)\epsilon,
\\ \nn &&W= 1+W_1 \epsilon,
\\ \nn &&v= v_0+(v_1+v_2\log\epsilon)\epsilon.\eea
Inserting into Eq.(\ref{C1}), we obtain
\bea\label{CLO1}  &&C_3^{\tilde{\eta}}\approx
\frac{16c_{\Delta}^{d}}{3\tilde{\eta}^2 \sqrt{\left(1+\frac{1}{\tilde{\eta}^2}\right)\chi_{p0}}}
\Bigg\{\sqrt{(1+\tilde{\eta}^2)\chi_{p0}}\ \mbox{arctanh}\frac{\tilde{\eta}\sqrt{\chi_{p0}}}{\sqrt{1+\tilde{\eta}^2}}
\\ \nn &&-\Bigg[\frac{W_1}{2} \sqrt{(1+\tilde{\eta}^2)\chi_{p0}}\ \mbox{arctanh}\frac{\tilde{\eta}\sqrt{\chi_{p0}}}{\sqrt{1+\tilde{\eta}^2}}
+\frac{\tilde{\eta}}{4\left(1+\tilde{\eta}^2(1-\chi_{p0})\right)}\times
\\ \nn &&\left((1+\tilde{\eta}^2)(\chi_{p0}-2\chi_{p1})-
4\left((1+\tilde{\eta}^2)\chi_{p0}+2W_1\left(1+\tilde{\eta}^2(1-\chi_{p0})\right)\right)
\log 2\right)\Bigg]\epsilon
\\ \nn &&-\frac{\tilde{\eta}}{4\left(1+\tilde{\eta}^2(1-\chi_{p0})\right)}
\left(\left((1+\tilde{\eta}^2)(\chi_{p0}-2\chi_{p2})
+2W_1\left(1+\tilde{\eta}^2(1-\chi_{p0})\right)\right)\right)\epsilon\log\epsilon\Bigg\}
.\eea

In view of Eqs.(\ref{param}) and (\ref{epsil}), we can express all the 
auxiliary parameters in terms of $v$ (or its coefficients 
$v_0$, $v_1$, and $v_2$):
\bea\label{chisol} &&\chi_{p 0}=1-v_0^2,\h \chi_{p 1}=1-v_0^2-2v_0 v_1-
\frac{(1-v_0^2)^2}{1+\tilde{\eta}^2 v_0^2},\h \chi_{p 2}=-2v_0 v_2,
\\ \nn &&W_1=-\frac{(1+\tilde{\eta}^2)(1-v_0^2)}{1+\tilde{\eta}^2 v_0^2}.\eea
This leads to
\bea\label{CLO2}  &&C_3^{\tilde{\eta}}\approx \frac{16 c_\Delta^d}{3 \tilde{\eta}}
\Bigg\{\mbox{arctanh}\frac{\tilde{\eta}\sqrt{1-v_0^2}}{\sqrt{1+\tilde{\eta}^2}}
+\frac{1}{4\sqrt{(1+\tilde{\eta}^2)(1-v_0^2)}{\left(1+\tilde{\eta}^2 v_0^2\right)^2}}\times
\\ \nn &&\Bigg[(1+\tilde{\eta}^2)\left((1-v_0^2)\left(1+\tilde{\eta}^2 v_0^2\right)
\left(2\sqrt{\left(1+\tilde{\eta}^2\right)((1-v_0^2)}\mbox{arctanh}\frac{\tilde{\eta}\sqrt{1-v_0^2}}{\sqrt{1+\tilde{\eta}^2}}
-\tilde{\eta}\log 16\right)\right.
\\ \nn &&\left. -\tilde{\eta} \left(1-v_0(3v_0-2v_0^3-4v_1+v_0(1-v_0^2-4v_0 v_1)\tilde{\eta}^2)\right)\right)\Bigg]\epsilon
\\ \nn &&+\frac{\tilde{\eta}(1+\tilde{\eta}^2)(1-v_0^2-4v_0 v_2)}
{4\sqrt{(1+\tilde{\eta}^2)(1-v_0^2)}(1+\tilde{\eta}^2 v_0^2)}\ \epsilon\log\epsilon\Bigg\}.\eea

To fix $v_0$, $v_1$, and $v_2$, 
one can use the small $\epsilon$ expansion of the angular difference
\bea\nn \Delta\phi_1=\phi_1(\tau,L)-\phi_1(\tau,-L)\equiv p,\eea
where we identified the angular difference
$\Delta\phi_1$ with the magnon momentum $p$ on the  worldsheet.
The result is \cite{AP2014}
\bea\label{v0sol} v_0=\frac{\cot\frac{p}{2}}
{\sqrt{\tilde{\eta}^2+\csc^2\frac{p}{2}}},\eea
and 
\bea\label{v1v2} v_1=\frac{v_0(1-v_0^2)\left[1-\log 16 +\tilde{\eta}^2
\left(2-v_0^2(1+\log 16 )\right)\right]}{4(1+\tilde{\eta}^2 v_0^2)},
\h v_2=\frac{1}{4}v_0(1-v_0^2).\eea
By using (\ref{v0sol}), (\ref{v1v2}) in (\ref{CLO2}), one finds
\bea\label{C3} &&C_3^{\tilde{\eta}}\approx \frac{16 c_\Delta^d}{3 \tilde{\eta}}
\Bigg\{\mbox{arcsinh}\left( \tilde{\eta} \sin\frac{p}{2}\right)+
\frac{(1+\tilde{\eta}^2)\sin^2\frac{p}{2}}{4\sqrt{\tilde{\eta}^2+\csc^2
\frac{p}{2}}}\times
\\ \nn &&\Bigg[\Bigg(2 \sqrt{\tilde{\eta}^2+\csc^2
\frac{p}{2}}\mbox{arcsinh}\left( \tilde{\eta} \sin\frac{p}{2}\right)
-\tilde{\eta}(1+\log 16)\Bigg)\epsilon +\tilde{\eta}\epsilon\log\epsilon\Bigg]
\Bigg\}.\eea

The expansion parameter $\epsilon$ in the leading order is given by
\cite{AP2014}
\bea\label{eps1} \epsilon =16\ \exp\left[-\left(\frac{J_1}{g}
+\frac{2\sqrt{1+\tilde{\eta}^2}}{\tilde{\eta}}\mbox{arcsinh}\left(\tilde{\eta} \sin\frac{p}{2}\right)
\right)\sqrt{\frac{1+\tilde{\eta}^2\sin^2\frac{p}{2}}{\left(1+\tilde{\eta}^2\right)\sin^2\frac{p}{2}}}
\right].\eea
Here we used Eq.({\ref{T}) for the string tension $T$.

The final expression for the normalized structure costant is given by
\bea\label{C3f} &&C_3^{\tilde{\eta}}\approx \frac{16 c_\Delta^d}{3 \tilde{\eta}}
\Bigg\{ \mbox{arcsinh}\left(\tilde{\eta}\sin\frac {p}{2}\right)-
4 \frac{\tilde{\eta}(1+\tilde{\eta}^2)\sin^3 \frac{p}{2}}{\sqrt{1+\tilde{\eta}^2
\sin^2\frac{p}{2}}}\Bigg[1+\frac{J_1}{g} \sqrt{\frac{\tilde{\eta}^2+\csc^2\frac{p}{2}}{1+\tilde{\eta}^2}}\Bigg]
\\ \nn &&\times \exp\left[-\left(\frac{J_1}{g}
+\frac{2\sqrt{1+\tilde{\eta}^2}}{\tilde{\eta}}\mbox{arcsinh}\left(\tilde{\eta} \sin\frac{p}{2}\right)
\right)\sqrt{\frac{1+\tilde{\eta}^2\sin^2\frac{p}{2}}{\left(1+\tilde{\eta}^2\right)\sin^2\frac{p}{2}}}
\right]\Bigg\}.\eea

Let us point out that in the limit $\tilde{\eta} \to 0$, (\ref{C3f}) reduces to
\bea\nn C_3\approx \frac{16}{3} c_\Delta^d \sin\frac{p}{2}
\left[1-4\sin\frac{p}{2}\left(\sin\frac{p}{2}+\frac{J_1}{g}\right)
\exp\left(-\frac{J_1}{g \sin\frac{p}{2}}-2\right)\right],\eea
which reproduces the result for the undeformed case found in \cite{AP2011}.
Another check is that this satisfies Eq.(\ref{check}) with
$\Delta$ computed in 
\cite{AP2014}
\bea\label{fr} &&\Delta\equiv E-J_1\approx 2 g \sqrt{1+\tilde{\eta}^2} \Bigg\{\frac{1}{\tilde{\eta}}
\mbox{arcsinh}\left(\tilde{\eta} \sin\frac{p}{2}\right)-4\frac{(1+\tilde{\eta}^2)
\sin^3\frac{p}{2}}{\sqrt{1+\tilde{\eta}^2 \sin^2\frac{p}{2}}}\times
\\ \nn && \exp\left[-\left(\frac{J_1}{g}
+\frac{2\sqrt{1+\tilde{\eta}^2}}{\tilde{\eta}}\mbox{arcsinh}\left(\tilde{\eta} \sin\frac{p}{2}\right)
\right)\sqrt{\frac{1+\tilde{\eta}^2\sin^2\frac{p}{2}}{\left(1+\tilde{\eta}^2\right)\sin^2\frac{p}{2}}}
\right] \Bigg\}.\eea

\setcounter{equation}{0}
\section{Concluding Remarks}
Here we obtained the {\it exact} semiclassical the 3-point correlation function between
two finite-size giant magnons ``heavy'' string states
and the ``light'' dilaton operator with zero momentum in the $\eta$-deformed $AdS_5\times S^5$.
It is given in terms of the complete elliptic integrals of first and third kind.
We proved the consistency of our result by taking a derivative of the conformal dimension w.r.t. the 
coupling constant.
We also provided the leading finite-size effect expansion of the structure constant.

It will be interesting to compute other three-point correlation functions of the $\eta$-deformed background
such as HHH to which our result may be useful.

\section*{Acknowledgements}
This work was partially supported by the Koran-Eastern European cooperation
in research and development through the National 
Research Foundation of Korea (NRF)
\\ (NRF-2013K1A3A1A39073412) (CA) and the Brain Pool program
(131S-1-3-0534) (PB) both funded by Ministry of Science, ICT 
and Future Planning.
CA thanks for the hospitality of String theory group in Humboldt University.

\vskip 1cm
\begin{appendix}
\leftline{{\bf Appendix:} The Mathematica code for the 
consistency check Eq.(\ref{check})}
\newcommand{\mathsym}[1]{{}}
\newcommand{\unicode}[1]{{}}

\newcounter{mathematicapage}
\vskip .5cm
\begin{doublespace}
\noindent\(\pmb{J[\text{g$\_$}]\text{:=}\frac{2 T}{\eta  \sqrt{(\text{xn}-\text{xm}[g]) \text{xp}[g]}}}\\
\pmb{\left(\text{xp}[g] \text{EllipticK}\left[\frac{(\text{xp}[g]-\text{xm}[g]) \text{xn}}{(\text{xn}-\text{xm}[g]) \text{xp}[g]}\right]-\text{xm}[g]
\text{EllipticPi}\left[1-\frac{\text{xm}[g]}{\text{xp}[g]},\frac{(\text{xp}[g]-\text{xm}[g]) \text{xn}}{(\text{xn}-\text{xm}[g]) \text{xp}[g]}\right]\right);}\\
\pmb{p[\text{g$\_$}]\text{:=}\frac{2 \text{xm}[g]}{\eta } \sqrt{\frac{1-\text{xp}[g] }{(1-\text{xm}[g]) (\text{xn}-\text{xm}[g]) \text{xp}[g]}}}\\
\pmb{\left(\text{EllipticK}\left[\frac{(\text{xp}[g]-\text{xm}[g]) \text{xn}}{(\text{xn}-\text{xm}[g]) \text{xp}[g]}\right]-\text{EllipticPi}\left[\frac{\text{xp}[g]-\text{xm}[g]}{(1-\text{xm}[g])
\text{xp}[g]},\frac{(\text{xp}[g]-\text{xm}[g]) \text{xn}}{(\text{xn}-\text{xm}[g]) \text{xp}[g]}\right]\right);}\\
\pmb{\text{En}[\text{g$\_$}]\text{:=}\frac{2 T (\text{xp}[g]-\text{xm}[g])}{\eta  \sqrt{(1-\text{xm}[g]) (\text{xn}-\text{xm}[g]) \text{xp}[g]}}
\text{EllipticK}\left[\frac{(\text{xp}[g]-\text{xm}[g]) \text{xn}}{(\text{xn}-\text{xm}[g]) \text{xp}[g]}\right];}\\
\pmb{T=\sqrt{1+\eta ^2} g;}\)
\end{doublespace}

\begin{doublespace}
\noindent\(\pmb{\text{Eq1}=D[J[g],g]==0;}\\
\pmb{\text{Eq2}=D[p[g],g]==0;}\\
\pmb{\text{sol}=\text{Solve}[\{\text{Eq1},\text{Eq2}\},\{D[\text{xm}[g],g],D[\text{xp}[g],g]\}];}\)
\end{doublespace}

\begin{doublespace}
\noindent\(\pmb{\text{xpd}=D[\text{xp}[g],g]\text{/.}\text{sol}[[1]];}\\
\pmb{\text{xmd}=D[\text{xm}[g],g]\text{/.}\text{sol}[[1]];}\)
\end{doublespace}

\begin{doublespace}
\noindent\(\pmb{\text{threept}=\frac{8 c}{3 \sqrt{1+\eta ^2}}\text{FullSimplify}[D[\text{En}[g]-J[g],g]\text{/.}\{D[\text{xp}[g],g]\to \text{xpd},D[\text{xm}[g],g]\to
\text{xmd}\}]}\)
\end{doublespace}

\begin{doublespace}
\noindent\(\frac{16 c \left(-\text{EllipticK}\left[\frac{\text{xn} (-\text{xm}[g]+\text{xp}[g])}{(\text{xn}-\text{xm}[g]) \text{xp}[g]}\right]+\text{EllipticPi}\left[1-\frac{\text{xm}[g]}{\text{xp}[g]},\frac{\text{xn}
(-\text{xm}[g]+\text{xp}[g])}{(\text{xn}-\text{xm}[g]) \text{xp}[g]}\right]\right) \text{xm}[g]}{3 \eta  \sqrt{(-1+\text{xm}[g]) (-\text{xn}+\text{xm}[g])
\text{xp}[g]}}\)
\end{doublespace}

\end{appendix}

\end{document}